\documentclass{article}
\usepackage{cite}
\usepackage{graphicx}

\begin{document}

\newcommand{\bear}{\begin{eqnarray}}
\newcommand{\eear}{\end{eqnarray}}
\newcommand{\be}{\begin{equation}}
\newcommand{\ee}{\end{equation}}
\newcommand{\beqn}{\begin{eqnarray}}
\newcommand{\eeqn}{\end{eqnarray}}
\newcommand{\beqnn}{\begin{eqnarray*}}
\newcommand{\eeqnn}{\end{eqnarray*}}

\newcommand{\exponente}[1]{\mathrm{e}^{#1}}
\newcommand{\adj}[1]{\left( #1 \right)}
\newcommand{\X}{\hat{x}}
\newcommand{\p}{\hat{p}}

\newcommand{\avg}[1]{\langle#1\rangle}

\def\vep{\varepsilon}
\def\vf{\varphi}
\def\al{\alpha}

\begin{center} {\Large \bf
Adiabatic amplification of energy and magnetic moment of a charged particle 
after the magnetic field inversion}

\end{center}

\begin{center} {\bf
Viktor V. Dodonov
and Alexandre V. Dodonov }
\end{center}
 
\begin{center}

{\it 
Institute of Physics,,
University of Brasilia, 
70910-900 Brasilia, Federal District, Brazil
\\
International Center for Physics, University of Brasilia, Brasilia, DF, Brazil }

\end{center}

{Correspondence: vdodonov@unb.br}

\abstract{
We study the evolution of the energy and magnetic moment of a quantum charged particle placed in a homogeneous
magnetic field, when this field changes adiabatically its sign.
We show that after a single magnetic field passage through zero value,
the famous adiabatic invariant ratio of energy to frequency  is reestablished again, 
but with the proportionality coefficient higher than in the initial state. 
The concrete value of this proportionality coefficient 
depends on the power index of the frequency dependence
on time near zero point. In particular, the adiabatic ratio of the initial ground state 
(with zero radial and angular quantum numbers)
triplicates if the frequency tends to zero linearly as function of time. 
If the Larmor frequency attains zero more than once, the adiabatic
proportionality coefficient strongly depends on the lengths of the time intervals between
zero points, so that the mean energy behavior can be quasi-stochastic after many passages
through zero value. 
The original Born-Fock adiabatic theorem does not work after the frequency passes 
through zero. However, its generalization is found: the initial Fock state becomes a wide
superposition of many instantaneous Fock states, whose weights do not depend on time in the new adiabatic regime.
}

\section{Introduction}
\label{sec-intr}

We consider the motion of a non-relativistic spinless particle of mass $M=1$ and charge $e$ in the $xy$-plane in the presence of
a uniform time-dependent magnetic field $B(t)$ directed along the $z$-axis (perpendicular to the plane).
Since this motion is independent from the motion along  the axis, it can be 
 described by means of the two-dimensional Hamiltonian
 (in the Gauss system of units)
\be
{H}(t) = \frac1{2}\sum_{j=1,2}\left[\hat{p}_j -{e}{ A}_j(t)/{c}\right]^2 .
\label{Ham1}
\ee
Here, $p_j$ and $A_j$ are components of the canonical momentum and vector potential,
respectively. 
In this paper, we assume the ``circular'' gauge of the vector potential, 
${\bf A}(t)=B(t)(-y,x)/2$.
Then,
\be
{H}(t) = \frac1{2}\left[\hat{p}_x^2  + \hat{p}_y^2 +\omega^2(t) \left(\hat{x}^2 +\hat{y}^2 \right) \right]
- \omega(t)\left(\hat{x}\hat{p}_y - \hat{y}\hat{p}_x \right), \qquad \omega(t) = eB(t)/(2c).
\label{Ham2}
\ee

Solutions of the stationary Schr\"odinger equation $\hat{H}\psi = E\psi$ with Hamiltonian (\ref{Ham2}) 
(including an additional 
isotropic harmonic oscillator potential) were found for the first time by Fock \cite{Fock28} and later
by Darwin \cite{Darwin31} in the special case of $B(t)=const$. 
These solutions can be written in polar coordinates as follows,
\be
\psi_{n_r m}(r,\vf) = \sqrt{\frac{\kappa n_r!}{\pi \left(n_r +|m|\right)!}}
\left(\kappa r^2\right)^{|m|/2}L_{n_r}^{(|m|)}\left(\kappa r^2\right)
 \exp\left(-{\kappa} r^2/2 +i m\vf\right), 
\label{psiFock}
\ee
\be
 \kappa = |\omega|/\hbar,
\qquad
m = 0, \pm 1, \pm 2, \ldots, \quad n_r =0,1,2, \ldots.
\ee
Function $L_n^{(\alpha)}(z)$ is the generalized Laguerre polynomial, defined 
as \cite{BE,Grad}
\[
 L_n^{(\alpha)}(z) = 
 \frac{1}{n!} e^z z^{-\alpha}\frac{d^n}{dz^n}\left(e^{-z} z^{\alpha +n}\right).
\]
The energy eigenvalues are given by the formula
\be
E_{n_r m} =\hbar|\omega|\left(1 +|m| +2n_r\right) -\hbar\omega m,  
\label{Emag}
\ee
which consists of two parts, in accordance with the two parts of Hamiltonian (\ref{Ham2}).
Note that the radial part of the wave function (\ref{psiFock}) depends on the absolute value $|m|$ of the canonical angular
momentum eigenvalue. The algebraic value $m$ (which can be either positive or negative) enters through the preserved
canonical angular momentum operator in Hamiltonian (\ref{Ham2}). 
If $\omega>0$, all states with $m\ge 0$ have the same energy $\hbar|\omega|\left(1  +2n_r\right)$, meaning an infinite
degeneracy of these energy levels.
On the other hand, the same infinite degeneracy happens for all states with $m \le 0$ if $\omega <0$.
These observations are important for the interpretation of results of the following sections.

What happens with the solution 
(\ref{psiFock}),  when the magnetic field depends on time after some initial time instant $t_i$?
A general answer was given many years ago by Malkin, Man'ko and Trifonov \cite{MMT70}: 
\be
\Psi_{n_r m}(r,\vf;t) = \sqrt{\frac{K(t) n_r! \left[K(t) r^2\right]^{|m|}}{\pi \left(n_r +|m|\right)!}}
L_{n_r}^{(|m|)}\left[K(t) r^2\right]
 \exp\left(i\frac{\dot\vep}{2\hbar\vep} r^2 +i m\vf + i \chi(t)\right). 
\label{psiFock-t}
\ee
The real function $ K(t) = \hbar^{-1}|\vep(t)|^{-2}$ is determined by
the complex function $\vep(t)$, which is the solution to the classical
equation of the harmonic oscillator with a time-dependent frequency,
\be
\ddot\vep +\omega^2(t)\vep =0,
\label{eqvep}
\ee
satisfying the initial conditions
\be
\vep(t_i)= [\omega(t_i)]^{-1/2}, \quad \dot\vep(t_i)= i[\omega(t_i)]^{1/2}.
\label{condvep}
\ee 
These conditions fix the value of the time-independent Wronskian
\be
\dot\vep \vep^* - \dot\vep^* \vep =2i.
\label{Wr}
\ee
We assume that $\omega(t_i) >0$.
Since the phase $\chi(t)$ is not important for our purposes, we do not bring here its explicit
(rather complicated) expression. 
Functions (\ref{psiFock-t}) are orthogonal and normalized:
\be
\langle \Psi_{n_r^{\prime} m^{\prime}}(t) | \Psi_{n_r m}(t)\rangle = \delta_{m m^{\prime}} \delta_{n_r^{\prime} n_r}.
\ee

Various results related to quantum systems described by means of time-dependent Hamiltonian 
(\ref{Ham2})
 were found for the past five decades after paper \cite{MMT70}: see, e.g., papers 
\cite{Aga80,Abdalla88a,Jan89,BasMiMo,DMMPR95,DelMi,Menouar10,ManZheb12,Aguiar16}.  
However, an interesting special case of time inversion of the magnetic field, especially in the adiabatic regime, was not 
considered in previous studies, in spite that the adiabatic approximation in quantum
mechanics was the subject of numerous papers: see, e.g., \cite{BF,MMT73,Marz04,Amin09,Yukalov09,Compar09,Rigol14,Weinberg17}.
 The aim of our paper is to fill in this gap (at least partially).
Remember that stationary homogeneous magnetic fields can be created inside solenoids with constant currents. On the other hand,
alternating currents are quite common in our daily lives. Hence, it can be interesting to understand, what can happen with initial 
quantum states of a charged particle placed inside a solenoid with an alternating current. It is worth noticing in this connection,
that periodic variations of the magnetic field created by a current 
with the standard frequency $50\,$Hz can be considered as adiabatic, if the amplitude of magnetic field is not too small.
Indeed,  
the Larmor frequency of an electron in the magnetic field of the order of $1\,$T
is of the order of $10^{11}\,$s$^{-1}$. This means that the period of rotation of the particle is many orders of magnitude smaller than the
scale of variations of the magnetic field. This presents one of the motivations behind our study.

The adiabatic evolution of the unidimensional harmonic oscillator, whose frequency passes through zero, was studied
in the recent paper \cite{aditrip}. However, the situation when the frequency turns into zero and returns to 
nonzero values for this system seems to be rather exceptional (although possible). On the other hand, a passage
of the Larmor frequency through zero is quite natural if the magnetic field changes its sign. In addition,
the existence of the second degree of freedom and additional invariant (the canonical angular momentum) adds more
interesting features to the dynamics of the system described by means of Hamiltonian (\ref{Ham2}).
One of these new features is the evolution of the magnetic moment. 
Another interesting feature is the infinite degeneracy of the stationary energy levels in the presence of magnetic field.
The analysis of these additional features is the second motivation of the present study.

The remainder of this paper is structured as follows.
In Section \ref{sec-adFock}, we bring general formulas describing the
adiabatic evolution of the Fock states and the magnetic moment, including the adiabatic
regimes without and with crossing zero value of the Larmor frequency.
Sections \ref{sec-upm} and \ref{sec-mult} are devoted to the analysis of the new adiabatic parameters
in the cases of a single and multiple transitions of the frequency through zero.
In Section \ref{sec-BF}, we consider the violation and generalization of the Born--Fock theorem
when magnetic field slowly changes its sign. Analytical results are illustrated with several figures.
In Section \ref{sec-meanen}, we study the evolution of the mean energy and the mean magnetic moment
(as well as their fluctuations) in the case of initial Fock states.
A more general case of initial  ``invariant states'' of the magnetic moment operator is considered in
Section \ref{sec-invarst}.
 Section \ref{sec-concl} contains the discussion of main results.

\section{Adiabatic evolution of the Fock states}
\label{sec-adFock}

The adiabatic (quasiclassical) approximate complex solution to Equation (\ref{eqvep}) with $\omega(t)>0$
can be written in the form
\be
\vep(t) \approx [\omega(t)]^{-1/2} e^{i\tilde\phi(t)}, 
\quad \dot\vep(t) \approx i[\omega(t)]^{1/2} e^{i\tilde\phi(t)}, 
\qquad \tilde\phi(t) = \int_{t_i}^t\omega(z)dz.
\label{adsol}
\ee
In this approximation, the solution (\ref{psiFock-t}) assumes the form (\ref{psiFock}) with the 
instantaneous value of function $\omega(t)$. This result agrees with the famous Born--Fock
adiabatic theorem \cite{BF}.
The mean energy evolves according to Equation (\ref{Emag}) with $\omega=\omega(t)$.
In particular, all states with $m\ge 0$ have the same mean energy $\hbar\omega(t)\left(1  +2n_r\right)$ if $\omega(t)>0$.

The magnetic moment operator has the form
\cite{JL49,Felder76,Friar81,March85,Stewart00,Bliokh12,Green15,Waka21,DHphys}
\be
\hat{\cal M} = \frac{e}{2c}\left( \hat{x}\hat{\pi}_y - \hat{y}\hat{\pi}_x\right) , \qquad 
\hat{\pi}_j \equiv \hat{p}_j - e A_j(\hat{x},\hat{y}).
\label{M2}
\ee
For the ``circular'' gauge of the vector potential, ${\bf A}=B(-y,x)/2$, we can write 
\be
\hat{\cal M} = \left[\hat{x} \hat{p}_y - \hat{y} \hat{p}_x - \omega \left(\hat{x}^2 + \hat{y}^2\right)\right]{e}/({2c}).
\label{M-Lam}
\ee
The magnetic moment operator does not commute with Hamiltonian (\ref{Ham2}):
\be
\left[\hat{H},\hat{M}\right] = i\omega\mu_B\left(\hat{x}\hat{\pi}_x 
+ \hat{\pi}_x \hat{x} + \hat{y}\hat{\pi}_y + \hat{\pi}_y \hat{y} \right),
\label{commHM}
\ee
where $\mu_B = e\hbar/(2c)$ is the Bohr magneton (for the unit mass chosen in this paper).

The mean value $\langle r^2\rangle \equiv \langle x^2 + y^2\rangle$ in the state (\ref{psiFock-t}) is given by a simple formula 
(see Appendix \ref{ap-intLag})
\be
\langle r^2\rangle_{n_r,m} = [K(t)]^{-1} \left(2n_r + |m| +1 \right).
\label{meanr2}
\ee
Consequently, the mean value of the magnetic moment in the Fock state (\ref{psiFock-t}) equals
\be
\langle{\cal M}\rangle_{n_r,m} = \mu_B\left[ m - \omega(t)|\vep(t)|^2 \left(2n_r + |m| +1 \right)\right].
\label{meanM-Fock}
\ee
This is not an eigenvalue of the magnetic moment operator, 
since the spectrum of this operator is continuous \cite{DD-pra20}.
Formula (\ref{meanM-Fock}) is interesting, because it shows that the {\em mean value\/} of the magnetic moment in the
energy eigenstate (when $\omega=const >0$ and $\omega|\vep|^2 =1$) is proportional to the energy eigenvalue, although
non-commuting operators $\hat{H}$ (\ref{Ham2}) and $\hat{M}$ (\ref{M-Lam}) have quite different structures and spectra.

Fluctuations of the magnetic moment, characterized by the variance
$\sigma_M = \langle \hat{\cal M}^2 \rangle - \langle \hat{\cal M} \rangle^2$, can be calculated
with the aid of formula (see Appendix \ref{ap-intLag})
\be
\langle r^4\rangle_{n_r,m} = [K(t)]^{-2} \left[ 6n_r\left(n_r + |m| +1 \right) +(|m|+1)(|m|+2)\right].
\label{meanr4}
\ee
The result is
\be
\sigma_M^{(n_r,m)} = \left[\mu_B \omega(t)|\vep(t)|^2\right]^2 \left[
2n_r\left(n_r + |m| +1 \right) +|m|+1\right].
\label{sigM}
\ee
We see that fluctuations of the magnetic moment can be strong in the initial energy eigenstates
(when $\omega|\vep|^2 =1$) with large values of the quantum numbers $n_r$ and $|m|$. But they are strong in the ground state
($n_r= m =0$) as well: $\sigma_M^{(0,0)} = [\langle{\cal M}\rangle_{0,0}]^2$.


In the adiabatic regime (\ref{adsol}), the mean magnetic moment does not depend on time, being an example of adiabatic invariants:
\be
 \langle{\cal M}\rangle_{n_r,m} = \mu_B\left[ m -  \left(2n_r + |m| +1 \right)\right] = 
- E_{n_r,m}(t)/[\hbar\omega(t)].
 \label{Madiab}
 \ee
However, the solution (\ref{adsol}) holds under the condition
\be
|\dot\omega|/\omega^2(t) \ll 1,
\label{cond}
\ee
which is obviously broken when $\omega(t)=0$.
Nonetheless, when the frequency slowly passes through zero value and slowly becomes not too small again, 
the conditions of the quasiclassical approximation are reestablished. 
We suppose that $\omega=0$ at some instant $t_* > t_i$. 
Then, the solution to Equation (\ref{eqvep}) 
can be written in the following most general quasiclassical form at $t \gg t_*$ 
[this formal relation means that time instant $t$ is so far from the instant $t_*$
that the condition (\ref{cond}) can be considered fulfilled]:
 \be
\vep(t) \approx |\omega(t)|^{-1/2} \left[ u_{+} e^{i{\phi}(t)} + 
 u_{-} e^{-i{\phi}(t)} \right], \qquad
 \phi(t) = \int_{t_*}^t |\omega(\tau)|d\tau,
\label{adsol+}
\ee
\be
\quad \dot\vep(t) \approx i|\omega(t)|^{1/2} 
\left[ u_{+} e^{i{\phi}(t)} -  u_{-} e^{-i{\phi}(t)} \right].
\label{deradsol+}
\ee
Constant complex coefficients $u_{\pm}$ must obey the condition  
\be
|u_{+}|^2 - |u_{-}|^2 =1,
\label{uvcond}
\ee
which is the consequence of Equation (\ref{Wr}). 
Pay attention to the choice of the lower limit of integration in the definition of the phase function $\phi(t)$.
This choice is insignificant for the solution in the form (\ref{adsol}), which is represented by a single exponential
function. But when one deals with a superposition of two exponential functions in (\ref{adsol+}), the choice of the
integration limits influences the phases of complex coefficients $u_{\pm}$. If the frequency passes through zero,
the time instant $t_*$ is distinguished. Therefore, the choice of $t_*$ as the starting point of integration seems
the most natural.
To fix the values of coefficients $u_{\pm}$, we choose the function $\vep(t)$ at $t \ll t_*$ in the form (\ref{adsol}),
but with the phase $\tilde\phi(t)$ replaced with the function $\phi(t)$ defined as in Equation (\ref{adsol+}).
Then, although $\phi(t) < 0$ for $t<t_*$, the time derivative $d\phi/dt = |\omega(t)|$ is positive.

In general, the mean magnetic moment oscillates at $t \gg t_*$:
\be
\langle{\cal M}\rangle_{n_r,m}
 = \mu_B\left[ m - \frac{\omega(t)}{|\omega(t)|} w(t)  \left(2n_r + |m| +1 \right)\right],
\label{meanM-Fock-neg}
\ee
\be
w(t) = |u_{+}|^2 + |u_{-}|^2 + 2\mbox{Re}\left[u_{+} u_{-}^* e^{2i{\phi}(t)} \right] .
\label{wt}
\ee
If $\omega(t)<0$, the function $w(t)$ enters Equation (\ref{wt}) with the positive sign. Otherwise, its contribution is negative. 
Due to Equation (\ref{uvcond}), we have the inequalities
\be
w_{min} =\left(\sqrt{1 + |u_{-}|^2} + |u_{-}|\right)^{-2} 
\le w(t) \le \left(\sqrt{1 + |u_{-}|^2} + |u_{-}|\right)^2 = w_{max}.
\label{wminmax}
\ee
Since $w_{min} <1$ and $w_{max} >1$ if $|u_{-}| >0$, 
the mean magnetic moment (\ref{meanM-Fock-neg}) oscillates between negative and positive
values for big enough (by absolute value) negative values of the quantum number $m$.

\section{Examples of adiabatic coefficients}
\label{sec-upm}

Exact values of the adiabatic coefficients are determined by the behavior of function $\omega(t)$ near zero value.
It was supposed in paper \cite{aditrip}, that absolute values of these coefficients 
can be expressed in terms of the exponent $n$
in the behavior $\omega^2(t) \sim |t-t_*|^n$ when $\omega(t)$ passes through zero. 
This hypothesis was based on the analysis of exact solutions to Equation (\ref{eqvep})
for the time-dependent frequency $\omega^2(t) =\omega_0^2 |t/\tau|^n$, when this equation
can be reduced to the Bessel equation. 
In that paper, devoted to the adiabatic dynamics of a quantum harmonic oscillator, 
the exponent $n$ could assume arbitrary non-negative values, as soon as function  $\omega^2(t)$ had physical meaning.
However, it is the function $\omega(t)$ that has physical meaning in the problems involving magnetic fields.
For this reason, supposing that function $\omega(t)$ can smoothly change its sign, we consider
here the special case of solutions obtained in \cite{aditrip}, when $n$ is an {\em even\/} number, i.e.,
\be
\omega(t) =\omega_0 (-t/\tau)^k, \qquad -\tau \le t \le \tau, \qquad k=0,1,2,\ldots.
\label{ompower}
\ee
Then, in the adiabatic limit $\omega_0\tau \gg 1$, the following formulas were derived in \cite{aditrip}:
\be
u_{+}^{-1} = \sin\left[\frac{\pi}{2(k+1)}\right], \qquad u_{-} = i\cot\left[\frac{\pi}{2(k+1)}\right].
\label{upmk}
\ee
Probably, the most natural situation takes place for $k=1$, i.e. the linear time dependence of a slow transition
of magnetic field through zero value. In this case, we have $|u_{-}|=1$ and $|u_{+}|^2=2$. 
Another interesting situation corresponds to $k=2$, when $|u_{-}|^2=3$. 
In this case, the magnetic field slowly diminishes to zero, 
but instead of crossing through zero, it slowly returns to the initial value.
For $k=3$ (a ``cubical'' crossing through zero frequency) we have $|u_{-}|^2= (\sqrt{2} + 1)^2$.

Another explicit exact solution to Equation (\ref{eqvep}) was found in paper \cite{D21} for the time-dependent Larmor frequency,
which changes continuously from the initial value $\omega_i$ to the final value $\omega_f$ as
\be
\omega(t) = \frac{\omega_f \exp(\kappa t) +\omega_i}{\exp(\kappa t) +1},
 \quad -\infty < t < \infty, \quad \kappa >0.
\label{omtanh}
\ee
Here, the positive parameter $\kappa$ characterizes the speed of frequency evolution. 
The following formula was obtained In the adiabatic regime, when $|\omega_i -\omega_f| \gg \kappa/2$:
\be
|u_-|^2 \approx \exp[2\pi(|\tilde\omega_i -\tilde\omega_f| -\tilde\omega_i -|\tilde\omega_f|)] , \qquad
\tilde\omega_{i,k} \equiv \omega_{i,k}/\kappa.
\label{u-2ad}
\ee
 If $\omega_i >\omega_f >0$, Equation (\ref{u-2ad}) yields 
$|u_-|^2 \approx \exp(-4\pi\tilde\omega_f) \ll 1$, so that the adiabatic invariance of the magnetic moment is preserved.
On the other hand, the same Equation (\ref{u-2ad}) yields $|u_-|^2 \approx 1$ for $\omega_f <0$ (provided $|\tilde\omega_f| \gg 1$).
We see that two different functions $\omega(t)$, given by Equation (\ref{ompower}) with $k=1$ and Equation (\ref{omtanh}), yield identical
{\em absolute values\/} of the adiabatic coefficients $|u_{\pm}|$. A common feature of these two functions is the {\em linear\/}
time dependence of the transition through zero value. 

However, the {\em phases\/} of adiabatic coefficients
are sensitive to the exact form of function $\omega(t)$. For example, it was shown in \cite{aditrip} that function (\ref{omtanh})
with $\omega_f = -\omega_i$
yields the complex adiabatic coefficients
\be
u_{+} = \sqrt{2}\, \exp\left[ -4i\tilde\omega_0 \ln(2)\right], \qquad
u_{-} = i.
\label{upm-tanh}
\ee
We see that phases of coefficients $u_{+}$ given by Equation (\ref{upmk}) with $k=1$ and Equation (\ref{upm-tanh})
are different.

\section{Multiple adiabatic passages of magnetic field through zero value}
\label{sec-mult}

What can happen if the Larmor frequency $\omega(t)$ passes adiabatically through zero value 
several times, at instants $t_0, t_1, t_2, \ldots$?
We suppose that we know all coefficients in the transition rules through zero frequency at each instant $t_k$:
\be
\left\{[\omega(t)]^{-1/2}e^{i\phi_k(t)}\right\}_{t \ll t_{k}} \to
\left\{[\omega(t)]^{-1/2}\left[u_{+}^{(k)}e^{i\phi_k(t)} + u_{-}^{(k)}e^{-i\phi_k(t)}\right]
\right\}_{t \gg t_{k}}, 
\label{trans}
\ee
\be
\phi_k(t) = \int_{t_k} ^t |\omega(z)|dz.
\ee
The resulting transformation after $N$ zero-crossings can be written as
\be
\left\{[\omega(t)]^{-1/2}e^{i\phi_0(t)}\right\}_{t \ll t_{0}} \to
\left\{[\omega(t)]^{-1/2}\left[U_{+}^{(N-1)}e^{i\phi_{0}(t)} + U_{-}^{(N-1)}e^{-i\phi_{0}(t)}\right]
\right\}_{t \gg t_{N-1}},
\label{transfin}
\ee
assuming that $U_{\pm}^{(0)} = u_{\pm}^{(0)}$. 
To establish the recurrence relations between the coefficients $U_{\pm}^{(N-1)}$ and $U_{\pm}^{(N)}$, 
we apply the rule (\ref{trans})
to the function arising after the $N$th zero,
using the superposition principle
and taking into account that the function $[\omega(t)]^{-1/2}e^{-i\phi_k(t)}$
 transforms as $\vep^*(t)$ after the frequency passes through zero value:
 \[
 \left\{[\omega(t)]^{-1/2}e^{-i\phi_k(t)}\right\}_{t \ll t_{k}} \to
\left\{[\omega(t)]^{-1/2}\left[u_{+}^{(k)*}e^{-i\phi_k(t)} + u_{-}^{(k)*}e^{i\phi_k(t)}\right]
\right\}_{t \gg t_{k}} .
\]
Then, using the relations
\be
\phi_0(t) = \Phi_k + \phi_k(t), \qquad \Phi_k = \int_{t_0}^{t_k} |\omega(z)|dz,
\ee
we can write the following equation for $t \gg t_N$:
\beqnn
U_{+}^{(N)}e^{i\phi_{0}(t)} + U_{-}^{(N)}e^{-i\phi_{0}(t)} &=& U_{+}^{(N-1)}e^{i\Phi_{N}}
\left[u_{+}^{(N)}e^{i\phi_N(t)} + u_{-}^{(N)}e^{-i\phi_N(t)}\right]
\\
&+& U_{-}^{(N-1)}e^{-i\Phi_{N}}
\left[u_{+}^{(N)*}e^{-i\phi_N(t)} + u_{-}^{(N)*}e^{i\phi_N(t)}\right].
\eeqnn
Equating the terms with the same time dependence $e^{\pm i\phi_N(t)}$, we arrive at the following recurrence relations:
\be
U_{+}^{(N)} = U_{+}^{(N-1)} u_{+}^{(N)} + U_{-}^{(N-1)} u_{-}^{(N)*} e^{-2i\Phi_{N}},
\ee
\be
U_{-}^{(N)} = U_{-}^{(N-1)} u_{+}^{(N)*} + U_{+}^{(N-1)} u_{-}^{(N)} e^{2i\Phi_{N}}.
\ee
One can verify that the identity $|U_{+}^{(N)}|^2 - |U_{-}^{(N)}|^2 =1$ is the consequence of identities 
\[
|U_{+}^{(N-1)}|^2 - |U_{-}^{(N-1)}|^2 =1, \qquad |u_{+}^{(N)}|^2 - |u_{-}^{(N)}|^2 =1.
\]
We see a strong dependence of coefficients $U_{\pm}^{(N)}$ on the phases $\Phi_{N}$, i.e., 
on the time intervals between the passages through zero frequency.
In the case of double passage through zero, the following inequalities hold:
\be
\left(|u_{+}^{(0)}u_{-}^{(1)}| - |u_{+}^{(1)}u_{-}^{(0)}|\right)^2 \le |U_{-}^{(1)}|^2 \le 
\left(|u_{+}^{(0)}u_{-}^{(1)}| + |u_{+}^{(1)}u_{-}^{(0)}|\right)^2.
\ee
In particular, two identical passages through zero frequency can result in the value $U_{-}^{(1)} =0$,
i.e., in reestablishing the standard adiabatic behavior, under the condition 
$\mbox{Re}\left[ u_{+}^{(0)}e^{i\Phi_1}\right] =0$.

\section{Generalization of the Born--Fock theorem}
\label{sec-BF}

The time-dependent solution (\ref{psiFock-t}) is a superposition of adiabatic eigenstates (\ref{psiFock})
with the instantaneous value of frequency $\omega(t)$ and the same value of the canonical
angular momentum quantum number $m$:
\be
\Psi_{n_r m}(r,\vf;t) = \sum_{q_r =0}^{\infty} C_{n_rq_r}^{(m)}(t) \psi_{q_r m}(r,\vf), \qquad
C_{n_rq_r}^{(m)}(t) = \langle \psi_{q_r m}|\Psi_{n_r m}(t) \rangle.
\label{expan}
\ee
Coefficients $C_{n_rq_r}^{(m)}(t)$ are oscillating functions of the  phase $\phi(t)$
[defined in Equation (\ref{adsol+})].
It is remarkable, however, that the modules squared $|C_{n_rq_r}^{(m)}|^2$ do not oscillate: they
depend on the absolute values of constant parameters $u_{\pm}$ only, as shown in Appendix \ref{ap-Jac}:
\be
|C_{n_rq_r}^{(m)}|^2 = \frac{\left(n_> +|m|\right)! n_<! |u_{-}|^{2|q_r - n_r|}}
{\left(n_< +|m|\right)! n_>! |u_{+}|^{2(|q_r - n_r| +|m|+1)}}
\left[ P_{n_<}^{(|q_r - n_r|, |m|)} \left(\frac{1-|u_{-}|^2}{1+|u_{-}|^2}\right)\right]^2.
\label{probab}
\ee
Here, $P_{k}^{(a,b)} (z)$ is the Jacobi polynomial. Other notations are as follows:
\[
n_< = \mbox{min}\left(q_r , n_r\right), \qquad n_> = \mbox{max}\left(q_r , n_r\right).
\]

Formula (\ref{probab}) has many interesting consequences. First, it does not depend on the sign of
quantum number $m$, while the energy levels (\ref{Emag}) are different for positive and negative
values of $m$. Second, the probabilities (\ref{probab}) are different for different positive values
of $m$, although the initial energy eigenvalues are the same, if $\omega(t<t_*) >0$.
If $\omega(t) < 0$ for $t>t_*$,  eigenstates $\psi_{q_r m}(r,\vf)$ in the expansion (\ref{expan}) describe 
the states with the instantaneous energy eigenvalues 
$E_{q_r m}(\omega<0) =\hbar|\omega(t)|\left(1 +|m| +m +2q_r\right)$,
which are different for different positive values of $m$, in contradistinction to the infinite energy levels degeneracy at $t< t_*$.

The simplest special case is $n_r =0$, when $n_<=0$ and $n_>=q_r$. 
Then, the Jacobi polynomial turns into unity, so that
\be
|C_{0q_r}^{(m)}|^2 = \frac{(q_r +|m|)! |u_{-}|^{2q_r}}{|m|! q_r! |u_{+}|^{2(q_r +|m| +1)}} .
\label{dist0}
\ee
One can verify that $\sum_{q_r=0}^{\infty} |C_{0q_r}^{(m)}|^2 =1$, due to the formula 
\be
S_m(x) =\sum_{k=0}^{\infty} \frac{(m + k)!}{m! k!} x^k = (1-x)^{-m-1}.
\label{Sm}
\ee

The mean energy in the distribution (\ref{dist0}) can be easily calculated with the aid of formula (\ref{Sm}):
\beqn
\langle \hat{H}(t) \rangle &=& \hbar|\omega(t)| \sum_{q_r=0}^{\infty} |C_{0q_r}^{(m)}|^2 
\left(1 +|m| -\frac{\omega(t)}{|\omega(t)|}m +2q_r\right)
\nonumber \\ 
&=& \hbar|\omega(t)|\left[1 +|m| -\frac{\omega(t)}{|\omega(t)|}m +2(1 +|m|)|u_{-}|^2 \right].
\label{meanE-nr0}
\eeqn
Remember that the initial energy was $E_{in} = \hbar|\omega_{in}|\left(1 +|m| -m\right)$,
and all initial states with $m\ge 0$ had the same energy. However, when the Larmor frequency
passes through zero and changes its sign, this infinite degeneracy is removed.
Moreover, the degeneracy is removed even when the frequency maintains its sign, if $u_{-} \neq 0$.
In any case, the adiabatic ratio $\langle \hat{H}(t) \rangle /|\omega(t)|$ 
{\em always\/} increases if the frequency passes through zero.

The distribution (\ref{dist0}) as function of $q_r$ has rather simple form. It decays monotonously
if $m=0$, going to a distribution with a single maximum at 
$q_r+1 \approx |m|\,|u_{-}|^2$ for large values of $|m|$. However,
the situation is more intricate for nonzero values of the initial radial quantum number $n_r =n$.
In the special case when $|u_{-}|^2 =1$ and $|u_{+}|^2 = 2$ (a single linear transition of the magnetic field through zero),
 the general distribution (\ref{probab}) assumes the form
\beqn
|C_{nq}^{(m)}|^2_{|u_{-}|=1} &=& \frac{\left(n_> +|m|\right)! n_<! }
{\left(n_< +|m|\right)! n_>! 2^{|q - n| +|m|+1}}
\left[ P_{n_<}^{(|q - n|, |m|)} (0)\right]^2
\nonumber \\ &=&
\frac{\left(n +|m|\right)! \left(q +|m|\right)! n! q!}{2^{n + q + |m| +1}}
\left[ \sum_{k=0}^{n_<} \frac{(-1)^k} {k! (n-k)! (q-k)! (k + |m|)!} \right]^2
.
\label{probab-u1}
\eeqn
\begin{figure}[hbt]
\vspace{-0.3cm}
\centering
\includegraphics[scale=0.2]{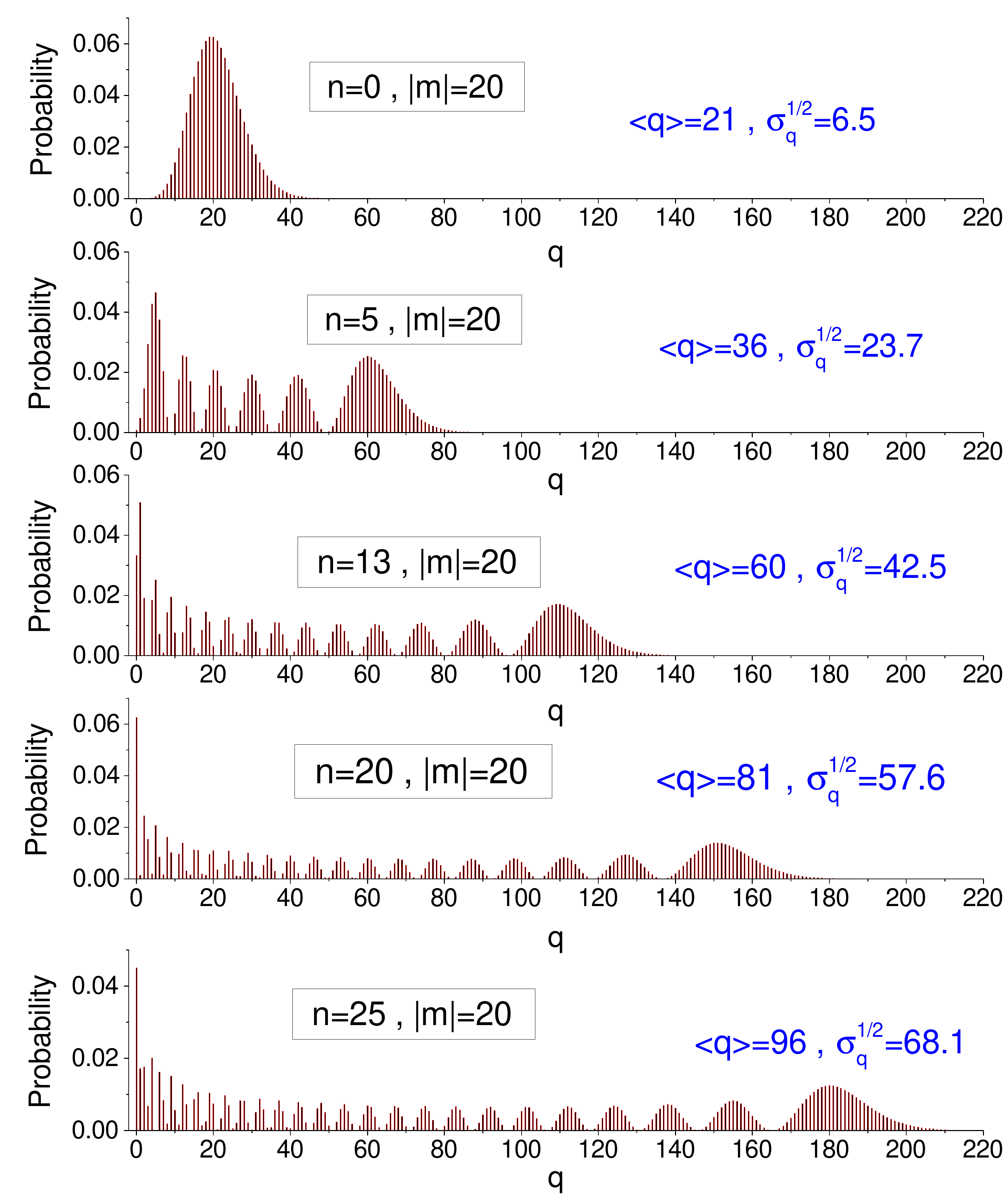}
\hspace{2cm}
\includegraphics[scale=0.2]{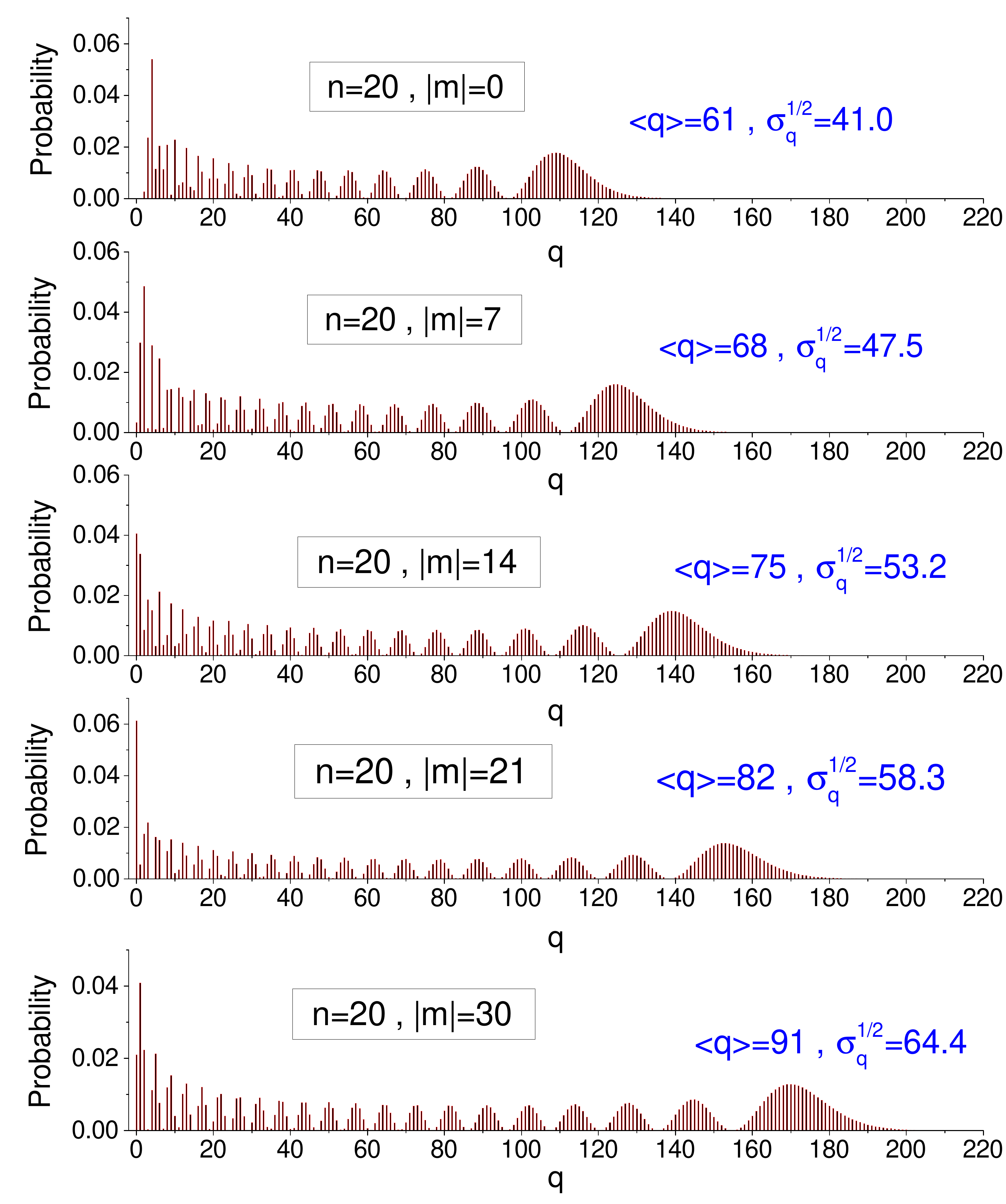}
\vspace{-0.3cm}
\caption{\small The probability distribution (\ref{probab-u1})
 of finding the initial Fock state $|n,m\rangle$ 
in the Fock state $|q,m\rangle$ 
after the frequency slowly passes through zero value, as function of $q$, in the case of 
$|u_{-}|=1$ and $|u_{+}|=\sqrt{2}$. Left: parameter $m$ is fixed. Right: parameter $n$ is fixed.
}
\label{fig-dist}
\end{figure} 
 
Several examples of the distribution (\ref{probab-u1}) as function of $q_r = q$ are shown
in Figure \ref{fig-dist} 
for various values of parameters $n_r =n$ and $|m|$.
We see that the distributions are very wide. Moreover, periodic structures are observed, and these structures
become almost regular for big values of number $q$. The amount of these structures is close to $n+1$ 
when parameters $n$ and $|m|$ are not too small. The symbols $\langle q\rangle$ and $\sigma_q^{1/2}$ stand for the average 
value of number $q$ and mean square deviation of this number, calculated numerically.

\begin{figure}[hbt]
\vspace{-2cm}
\centering
\includegraphics[scale=0.2]{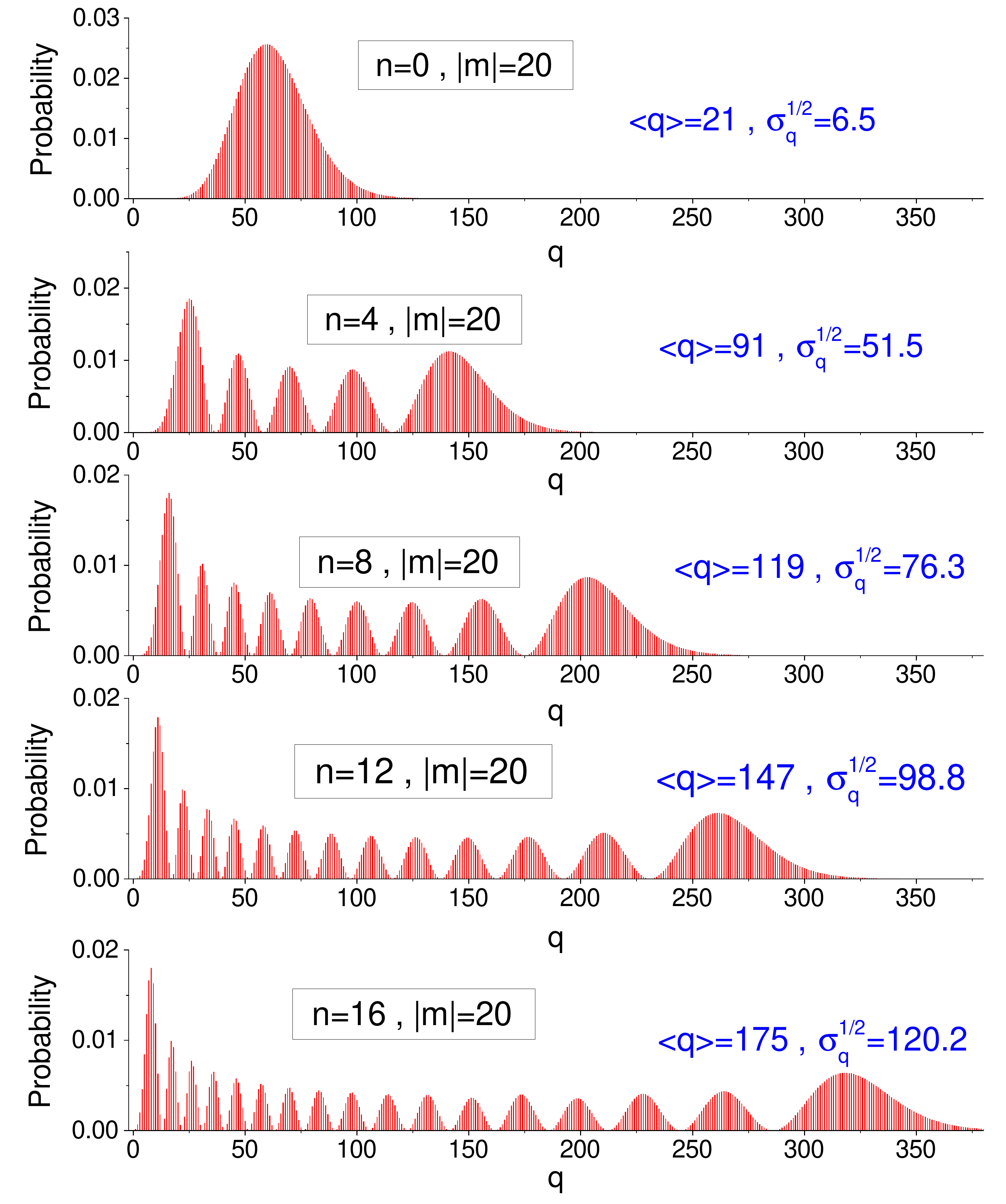}
\hspace{2cm}
\includegraphics[scale=0.2]{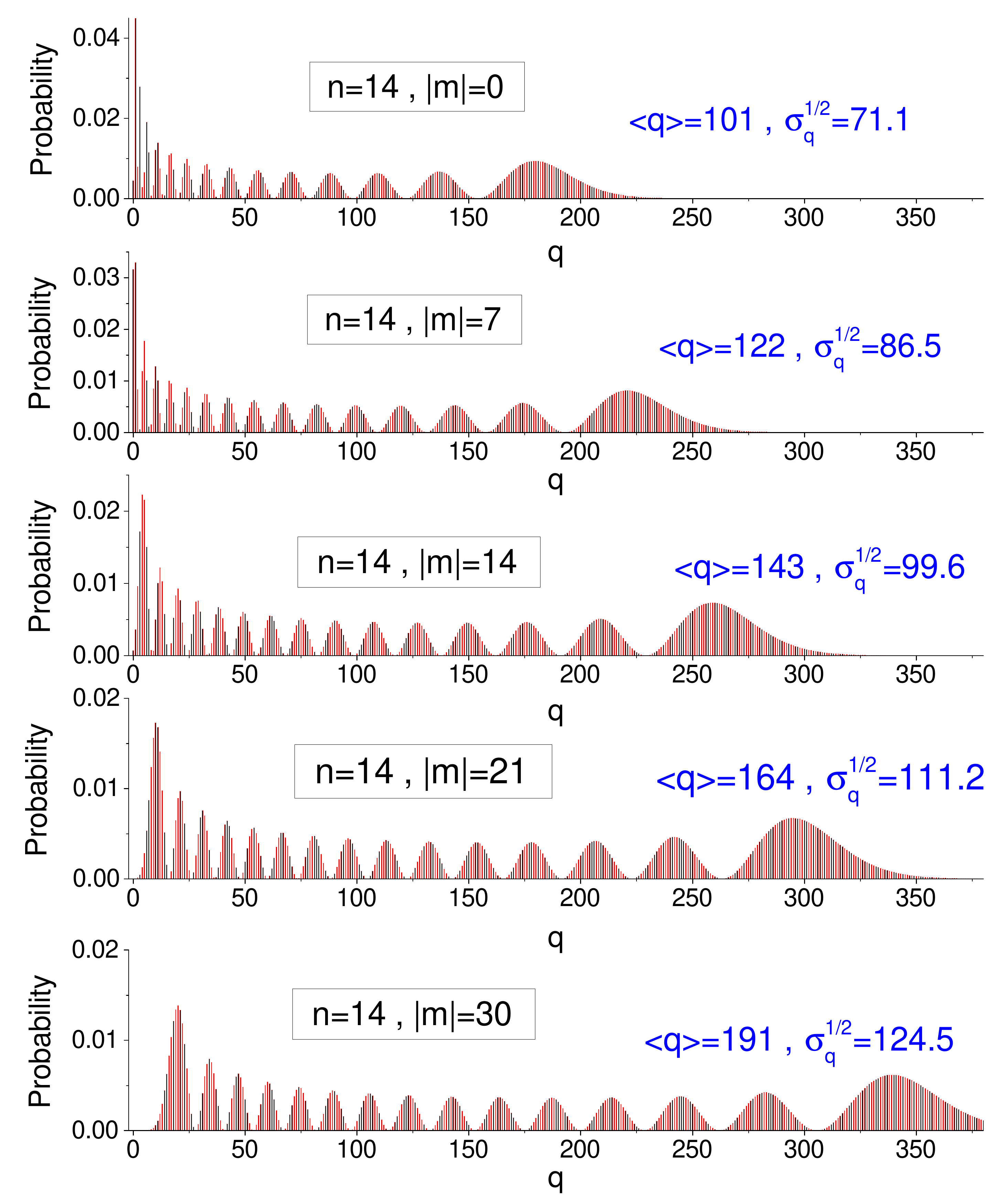}
\vspace{-0.3cm}
\caption{\small The probability distribution (\ref{probab-u1})
 of finding the initial Fock state $|n,m\rangle$ 
in the Fock state $|q,m\rangle$ 
after the frequency slowly passes through zero value, as function of $q$, in the case of 
$|u_{-}|=\sqrt{3}$ and $|u_{+}|={2}$. Left: parameter $m$ is fixed. Right: parameter $n$ is fixed.
}
\label{fig-dist+}
\end{figure}

\begin{figure}[hbt]
\centering
\includegraphics[scale=0.18]{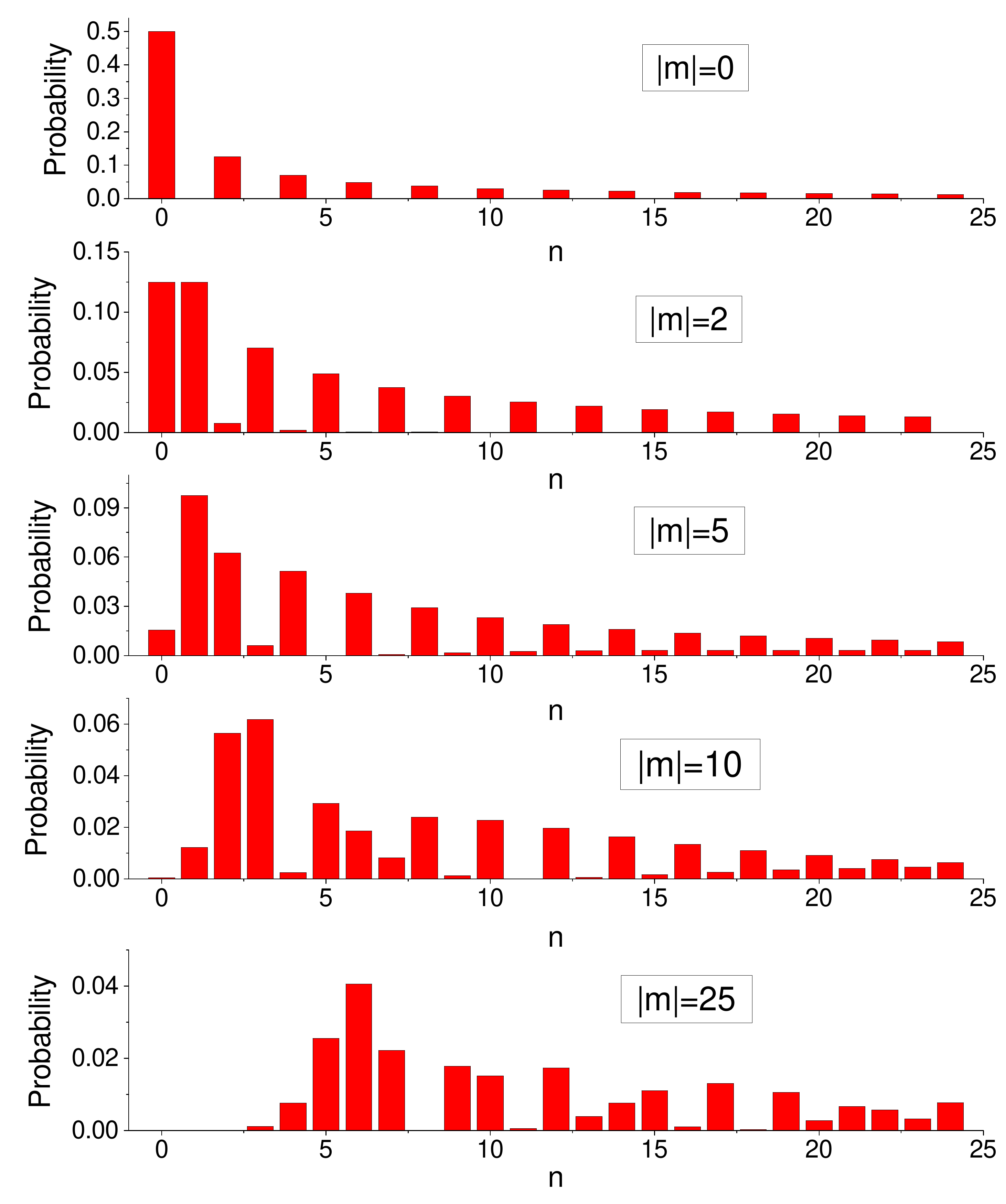}
\hspace{2cm}
\includegraphics[scale=0.18]{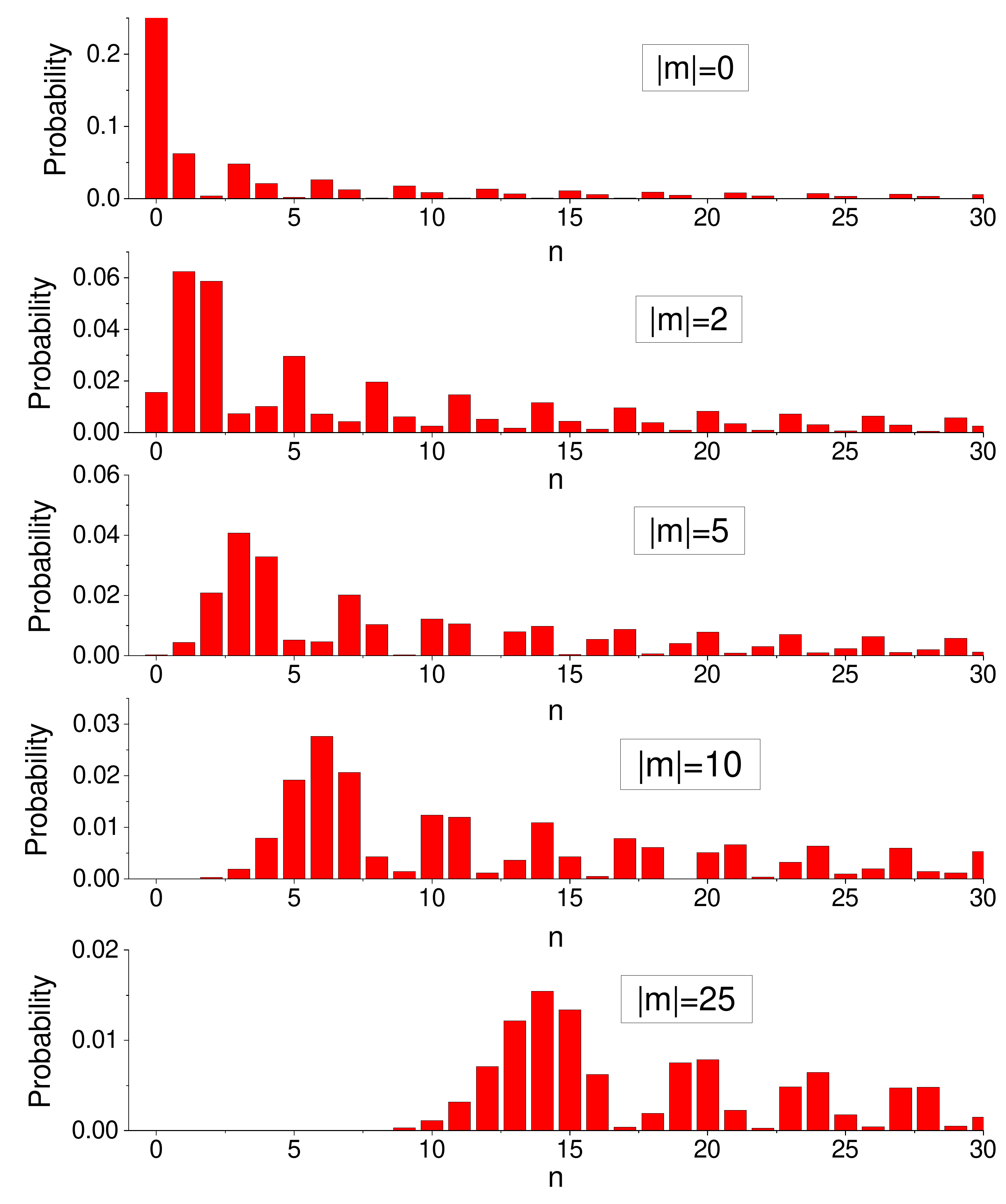}
\vspace{-0.3cm}
\caption{\small The probability 
 of finding the initial Fock state $|n,m\rangle$ 
in the same Fock state 
after the frequency slowly passes through zero value, as function of $n$ for different fixed values of the angular moment quantum number
$|m|$. Left: for $|u_{-}|=1$ and $|u_{+}|=\sqrt{2}$. Right: for $|u_{-}|=\sqrt{3}$ and $|u_{+}|={2}$.
}
\label{fig-nn}
\end{figure}

In Figure \ref{fig-dist+}, we show similar distributions for $|u_{-}|=\sqrt{3}$ and $|u_{+}|={2}$.
These values correspond to the ``parabolic'' form of function $\omega(t)$ near the point $t_*$, when the
Larmor frequency attains zero, but it does not change its sign, returning slowly to positive values.
In this case,
\beqn
|C_{nq}^{(m)}|^2_{|u_{+}|=2} &=& \frac{\left(n_> +|m|\right)! n_<! (3/4)^{|q - n|}}
{\left(n_< +|m|\right)! n_>! 4^{|m|+1}}
\left[ P_{n_<}^{(|q - n|, |m|)} (-1/2)\right]^2
\nonumber \\ &=&
\frac{\left(n +|m|\right)! \left(q +|m|\right)! n! q! 3^{q + n}}{4^{ |m| +1 + q + n}}
\left[ \sum_{k=0}^{n_<} \frac{(-3)^{-k}} {k! (n-k)! (q-k)! (k + |m|)!} \right]^2
.
\label{probab-u1+}
\eeqn
The plots of distribution (\ref{probab-u1+}) are similar to those for distribution (\ref{probab-u1}),
with the same number $n_r +1$ of periodic structures, but with an increased mean value 
$\langle q\rangle$. A simple analytic formula for this mean value is derived in Section \ref{sec-meanen}.
The probability $|C_{nn}^{(m)}|^2$ of remaining in the initial Fock state turns out very low for many 
initial quantum numbers $n$, 
except for the case of $n=m=0$: see Figure \ref{fig-nn}.
%

\section{Mean energy}
\label{sec-meanen}

To find the mean energy after the inversion of the magnetic field for an arbitrary initial energy
eigenstate, one could try to calculate the sum 
\be
\sum_{q_r=0}^{\infty} |C_{n_rq_r}^{(m)}|^2 \left(1 +|m| -\frac{\omega(t)}{|\omega(t)|}m +2q_r\right)
= 
1 +|m| -\frac{\omega(t)}{|\omega(t)|}m + 2\sum_{q_r=0}^{\infty} |C_{n_rq_r}^{(m)}|^2q_r 
.
\label{sumqr}
\ee
However, this way is not easy, because the sum contains squares of the Jacobi polynomials
with variable lower and upper indexes.
A direct calculation of the kinetic energy mean value $\langle \hat{p}_x^2 + \hat{p}_y^2\rangle/2$
requires a knowledge of complicated integrals containing derivatives of function (\ref{psiFock-t}).
Probably, the simplest way is to notice that Hamiltonian (\ref{Ham2}) is quadratic with respect to
the coordinates and momenta operators. Consequently, the Heisenberg or Ehrenfest equations for
these operators are linear. This means that their time-dependent mean values are linear combinations
of the initial values with certain time-dependent coefficients. 
It is convenient to introduce  vector ${\bf Q} = (x,y,p_x,p_y)$ (whose components are either mean values of quantum operators
or classical variables). Then, 
\be
{\bf Q}(t) = \Lambda(t;t_i){\bf Q}(t_i),
\label{LamQ}
\ee
where $\Lambda(t;t_i)$ is a $4\times4$ matrix.
Combining mean values  $Q_{jk} \equiv \langle \hat{Q}_j\hat{Q}_k + \hat{Q}_k\hat{Q}_j\rangle/2$ into the 
$4\times4$ symmetric matrix ${\cal Q}$, one can verify the relation (see \cite{183vol})
\be
{\cal Q}(t)= \Lambda(t;t_i) {\cal Q}(t_i) \tilde{\Lambda}(t;t_i),
\label{sigLamsig}
\ee
 where $\tilde{\Lambda}$ means the transposed matrix. 
In our case, matrix $\Lambda(t;t_i)$ can be expressed in terms of the solution $\vep(t)$ and its derivative
$\dot\vep(t)$ as follows \cite{D21},
\be
{\Lambda}(t;t_i) = \omega_{i}^{1/2} \left\Vert
\begin{array}{c c}
 \mbox{Re}(\vep) {\bf R} & \mbox{Im}(\vep) {\bf R}/\omega_{i} \\
 \mbox{Re}(\dot\vep) {\bf R} & \mbox{Im}(\dot\vep) {\bf R}/\omega_{i} 
\end{array}
\right\Vert, \qquad 
{\bf R} = \left\Vert
\begin{array}{c c }
\cos\tilde\phi	 &	\sin\tilde\phi	\\ 
-\sin\tilde\phi & \cos\tilde\phi 
\end{array}
\right\Vert ,
\label{LamR}
\ee 
with the phase $\tilde\phi$ defined in Equation (\ref{adsol}).
Comparing Equations (\ref{Ham2}), (\ref{Emag}) and (\ref{meanr2}), we obtain the initial mean value
\[
\langle \hat{p}_x^2 + \hat{p}_y^2\rangle_{n_r m} = \omega_i \left(2n_r + |m| +1 \right).
\]
Moreover, since the initial wave function (\ref{psiFock}) possesses the rotational symmetry, 
the initial covariance matrix ${\cal Q}(t_i)$ can be written in the following block form:
\be
{\cal Q}(t_i) = \frac{\hbar}{2}\left\Vert
\begin{array}{c c }
I\gamma/\omega_i & m\Sigma \\
-m\Sigma & I \gamma\omega_i
\end{array}
\right\Vert,
\;\; 
I = \left\Vert
\begin{array}{c c }
1 & 0 \\
0 & 1
\end{array}
\right\Vert,
\;\; 
\Sigma =  \left\Vert
\begin{array}{c c }
0 & 1 \\
-1 & 0
\end{array}
\right\Vert,
\label{Q0XK}
\ee 
where $\gamma = 2n_r + |m| +1$.
In this case, we obtain
\be
{\cal Q}(t) = \frac{\hbar}{2}\left\Vert
\begin{array}{c c }
I\gamma |\vep(t)|^2 & m\Sigma  + I \gamma \mbox{Re}\left(\dot\vep \vep^*\right) \\
-m\Sigma + I \gamma \mbox{Re}\left(\dot\vep \vep^*\right) & I \gamma|\dot\vep(t)|^2
\end{array}
\right\Vert.
\label{Qt}
\ee
Consequently,
\be
\langle \hat{H}(t) \rangle = \frac{\hbar\gamma}{2}\left[ |\dot\vep(t)|^2 + |\vep(t)|^2 \omega^2(t)
\right] - m \hbar \omega(t).
\label{Etvep}
\ee
In the adiabatic regime, 
using Equation (\ref{adsol+}), we obtain the formula
\be
\langle \hat{H}(t) \rangle = \hbar |\omega(t)| \left(2n_r + |m| +1\right) \left(|u_{+}|^2 + |u_{-}|^2 \right)
-\hbar\omega(t) m ,
\label{minE-ad-gen}
\ee
which goes to (\ref{meanE-nr0}) if $n_r =0$.
Comparing Equations (\ref{meanM-Fock-neg}) and (\ref{wt}) with Equation (\ref{minE-ad-gen}), 
we see that Equation
(\ref{Madiab}) can be generalized 
as follows,
\be
 \overline{\langle{\cal M}\rangle} = - \langle \hat{H}(t) \rangle/[\hbar\omega(t)],
 \label{Madiabmed}
 \ee
where the overline means an additional time averaging over fast oscillations of the mean magnetic
moment, as soon as $\overline{w(t)} = |u_{+}|^2 + |u_{-}|^2 $.

Comparing Equations (\ref{sumqr}) and (\ref{minE-ad-gen}), it is easy to see the equality
\[
\left(1 + 2|u_{-}|^2\right)\left(2n_{r} + |m| +1\right) = 2\langle q \rangle + |m| +1,
\]
which results in the formula
\be 
\langle q \rangle = n_{r}\left(1 + 2|u_{-}|^2\right) + |u_{-}|^2 (|m| +1).
\label{meanq}
\ee
In particular, we have $\langle q \rangle = 3 n_{r} + |m| +1$ for $|u_{-}|^2=1$ and
$\langle q \rangle = 7 n_{r} + 3(|m| +1)$ for $|u_{-}|^2=3$.
These relations coincide with the results of numeric calculations given in Figures 
\ref{fig-dist}--\ref{fig-dist++}.

It is interesting to calculate the variance 
$\sigma_H = \langle \hat{H}^2 \rangle - \langle \hat{H} \rangle^2$, characterizing the energy fluctuations.
This variance equals zero in the initial Fock state. However, it becomes nonzero when the frequency passes through zero.
Relatively simple calculations can be performed if $n_r =0$. Then, the sum
\[
\sum_{q_r=0}^{\infty} |C_{0q_r}^{(m)}|^2 
\left(1 +|m| -\frac{\omega(t)}{|\omega(t)|}m +2q_r\right)^2  
\] 
can be found with the aid of formula (\ref{Sm}) and its consequences.
The results are as follows,
\beqnn
\langle \hat{H}^2 \rangle &=& [\hbar\omega(t)]^2 \left[ \left(1 -\frac{\omega(t)}{|\omega(t)|}m +|m| \right)^2 
+4 (1+|m|)\left(2 -\frac{\omega(t)}{|\omega(t)|}m +|m| \right)|u_{-}|^2 \right. \\ && \left.  
+4 (1+|m|)(2+|m|)|u_{-}|^4\right],
\eeqnn
\be
\sigma_H = 4[\hbar\omega(t)]^2 (1+|m|)|u_{+}u_{-}|^2.
\label{sigH}
\ee
Note that the variance (\ref{sigH}) does not depend on the sign of quantum number $m$, while the mean value 
(\ref{minE-ad-gen}) is sensitive to this sign.
If $n_r >0$, the calculations become rather cumbersome, so we do not perform them here.
For explicit expressions in some special cases, one can consult papers \cite{DHphys,DH21}.

\section{Adiabatic evolution of the ``invariant states'' of the magnetic moment operator}
\label{sec-invarst}

The stationary Fock states (\ref{psiFock}) are determined by integral parameters $n_r = 0,1,2,\ldots$ and
$m=0,\pm 1,\pm2, \ldots$. In these normalized states, the energy and canonical angular momentum have definite values. 
In addition, the {\em mean value\/} of the magnetic moment operator (\ref{M-Lam}) does not depend on time in these states.
Recently, a wider family of {\em isotropic\/} states, possessing time-independent mean values of the magnetic moment operator
for the time-independent Hamiltonian (\ref{Ham2})
and named as ``magnetic moment invariant states'',
was found in paper \cite{D-invmag}. They are determined by two {\em continuous\/} positive parameters, $G_+ \ge 1$ and $G_- \ge 1$,
with the following nonzero mean values (provided $\omega >0$):
\be
\langle{\pi_x^2}\rangle = \langle{\pi_y^2}\rangle =  \langle \hat{H}\rangle = \hbar \omega G_{+},
\label{meanHG+}
\ee
\be
\langle \hat{x}\hat{p}_y\rangle = - \langle\hat{y}\hat{p}_x \rangle = \frac{\hbar}{4}\left(G_{-} - G_{+}\right),
\qquad \langle{y\pi_x}\rangle = - \langle{ x\pi_y}\rangle = \frac{\hbar}{2} G_{+} , 
\label{LcanH}
\ee
\be
\langle{p_x^2}\rangle = \langle{p_y^2}\rangle = \omega^2 \langle{x^2}\rangle
 = \omega^2\langle{y^2}\rangle = \frac14{\hbar\omega}\left( G_{+} + G_{-}\right).
\label{varGfree}
\ee
The solution (\ref{psiFock}) results in formulas (\ref{meanHG+})-(\ref{varGfree})
 with the {\em integral\/} positive coefficients
\be
G_{\pm} = 1 + 2 n_r + |m|  \mp {m}  . 
\label{Gint}
\ee

It is known that there exist many quite different quantum states with the same mean values of canonical operators and
their powers (products). In particular, one can construct  a two-parameter family of {\em Gaussian\/} states possessing 
the second-order moments (\ref{meanHG+})-(\ref{varGfree}) with arbitrary coefficients $G_{\pm}\ge 1$.
Their Wigner functions have the following form \cite{D-invmag} [here ${\bf r} = (x,y)$ and ${\bf p} = (p_x,p_y)$]:
\be
W_G({\bf r},{\bf p}) = \frac{4}{G_{+}G_{-}} \exp\left\{-\,\frac{(G_{+}+G_{-})\left[\omega^2{\bf r}^2 +{\bf p}^2\right] +
2(G_{+} - G_{-})\omega(xp_y - yp_x)}{2\hbar\omega G_{+}G_{-}}
\right\}.
\label{Wsymm}
\ee
The quantum purity of the state (\ref{Wsymm}) equals
\be
{\cal P} = \int W^2({\bf r},{\bf p})\frac{ d{\bf r} d{\bf p}}{(2\pi\hbar)^2} =
 ({G_{+}G_{-}})^{-1}.
\ee
Consequently, all Gaussian ``magnetic moment invariant states'' are {\em mixed\/} if $G_{\pm} > 1$, in contradistinction to the
{\em pure\/} quantum states (\ref{psiFock}) with integral values of parameters $G_{\pm}$, given by Equation (\ref{Gint}).
The meaning of parameters $G_{+}$ and $G_{-}$ becomes more clear, if one goes from the canonical operators in the phase space
to the relative $(x_r,y_r)$ and guiding center $(x_c,y_c)$ coordinates, 
\be
x_r = -\pi_y/(2m\omega), \quad y_r = \pi_x/(2m\omega), \qquad
x_c = x + \pi_y/(2m\omega), \quad y_c = y - \pi_x/(2m\omega).
\ee
The importance of these integrals of motion was emphasized by many authors during decades 
\cite{Land30,JL49,Dulock66,MM69,FeKa70,Tam71,32,33,34,DMP94,Li99,Kowalski05,Gazeau09,Mielnik11,Dodcohmag18,Champel19,Waka20,Kita,Fletcher21}.
Equivalent integrals of motion, obtained by the multiplication of $x_c$ and $y_c$ by $m\omega$, were considered under
the name ``pseudomomentum'' in papers \cite{32,Konst16,vanEnk20}.

The corresponding second-order moments are as follows,
\be
 \langle{x_r^2}\rangle = \langle{y_r^2}\rangle = \frac{\hbar}{4\omega} G_{+}, \qquad
  \langle{x_c^2}\rangle = \langle{y_c^2}\rangle = \frac{\hbar}{4\omega} G_{-},
\label{varxrxc}
\ee
with zero mean values of all cross-products.
Now, the inequalities $G_{\pm} \ge 1$ follow from the commutators 
$\left[\hat{x}_r, \hat{y}_r\right] =  \left[\hat{y}_c, \hat{x}_c\right] = i\hbar/(2\omega)$
and the  Heisenberg--Weyl (actually, Robertson's \cite{Robertson29}) uncertainty relation 
\be
 \langle{A^2}\rangle\langle{B^2}\rangle \ge |\langle [\hat{A},\hat{B}]\rangle|^2/4.
 \ee

The evolution of the second-order moments can be calculated in the same way as it was shown in Section \ref{sec-meanen}.
It is sufficient to replace $\gamma$ with $\left(G_{+} + G_{-}\right)/2$ and $m$ with $\left(G_{-} - G_{+}\right)/2$
in Equations (\ref{Q0XK}) and (\ref{Qt}). Hence, the mean values of energy and magnetic moment for $t>t_i$
can be written as follows,
\be
{\cal E}(t) \equiv \langle \hat{H}(t)\rangle = \frac{\hbar}{4}(G_{+} + G_{-})
\left[ |\dot\vep(t)|^2 + \omega^2(t)|\vep(t)|^2\right] +\frac{\hbar}{2}\omega(t)(G_{+} - G_{-}),
\label{EGt}
\ee
\be
{\cal M}(t) =\frac{\mu_B}{2} \left[ G_{-} - G_{+} - \omega(t)|\vep(t)|^2 (G_{+} + G_{-})\right].
\label{MGt}
\ee

If the frequency slowly goes to a constant asymptotic value $\omega_f$ (which can be positive or negative), then
the quantum mechanical mean value of the energy tends to a time-independent value
\be
{\cal E}_{f}  
= \frac{\hbar}{2}\left[|\omega_f|\left(G_{+} + G_{-}\right)\left(1 +2 |u_{-}|^2\right) + 
\omega_{f}\left(G_{+} - G_{-}\right) \right].
\label{Ef}
\ee
The quantum-mechanical mean value of the kinetic angular momentum performs  harmonic
oscillations with frequency $2|\omega_f|$ between the values ${\cal M}_{min}$ and ${\cal M}_{max}$.
The amplitude of these oscillations equals
\be
\Delta{\cal M} = \frac12\left({\cal M}_{max} -{\cal M}_{min}\right) 
=\mu_B |u_{+}u_{-}| \left(G_{+} + G_{-}\right).
\ee
The time-averaged value of these oscillations equals
\be
\overline{\langle {\cal M}\rangle} 
= \frac{\mu_B}{2}\left[ G_{-} - G_{+}
- \frac{\omega_f}{|\omega_f|}\left(G_{+} + G_{-}\right) \left( 1 + 2 |u_{-}|^2\right) \right]
.
\label{<<L>>}
\ee


\section{Conclusions}
\label{sec-concl}

The main result of the paper is the discovery of existence of generalized 
adiabatic invariants and a generalized Born--Fock adiabatic theorem for a charged particle
in a homogeneous magnetic field, when this field slowly passes through zero value.
According to the original Born--Fock theorem, the initial discrete energy eigenstates 
maintain their forms during the adiabatic evolution of the Hamiltonian, so that parameters
of the wave functions correspond to instantaneous values of coefficients of the Hamiltonian.
This statement becomes invalid when the time-dependent Larmor frequency approaches zero value.
We considered the situation when the frequency slowly goes away from zero, in such a way that
the adiabatic condition (\ref{cond}) is reestablished. Then, the evolved quantum state becomes
a {\em superposition\/} of instantaneous energy eigenstates with different radial quantum numbers
(but the same angular quantum numbers, due to the conservation of the canonical angular momentum). 
The generalized Born--Fock theorem
is the statement that the weight of each member of this superposition does not depend on time, 
as soon as the condition (\ref{cond}) is fulfilled. It is remarkable that all these weights
depend on the single parameter $|u_{-}|$ of the most general adiabatic solution (\ref{adsol+})
to the classical equation (\ref{eqvep}) of the harmonic oscillator with a time-dependent frequency.
Using examples of exact solutions, we believe (although we have no rigorous proof) that
the value of $|u_{-}|$ in the case of a single transition of frequency through zero is determined
by the exponent $k$ in the frequency behavior near zero: $\omega(t) \sim (t-t_*)^k$.
In particular, $|u_{-}| =1$ if $k=1$.
In the cases of multiple frequency passages through zero, the final coefficient $|u_{-}|$ 
is very sensitive to the additional parameter - the phase $\Phi$, which is the integral
of the absolute value $|\omega(t)|$ between the first and last zero frequency instants.
As a consequence, the adiabatic behavior after
many crossings through zero frequency can be quasi-chaotic.
Under some specific conditions, the mean energy can even return to the
initial value after multiple frequency passage through zero, while it can be significantly 
amplified under other specific condition. However, in all the cases, the adiabatic ratio
$\langle \hat{H}(t)\rangle/[\hbar|\omega(t)]$ can only increase after many frequency passages
through zero value. This ratio coincides (by the absolute value) with the double mean value
of the magnetic moment in the adiabatic regime. The word ``double'' means the quantum-mechanical
averaging accompanied with the time averaging over fast oscillations in time.

Note that the concepts of adiabaticity and multiple passages through zero frequency are totally
compatible. For example, the Larmor frequency of electron in the magnetic field of the order of $1\,$T
is of the order of $10^{11}\,$s$^{-1}$. In such a case, periodic variations of the magnetic field
with the standard frequency $50\,$Hz are totally adiabatic. It would be interesting to see, 
what can happen with the statistical properties of the energy and magnetic moment 
in this realistic situation
after a great number (of an order of $10^2$ or higher) of the frequency transitions through zero. 
But this can be a subject of another study. There are several other challenging problems.
In particular, in the case of very slow evolution, various {\em dissipation\/} effects can change
the evolution drastically. Another problem is to understand the adiabatic evolution for different gauges
of the vector potential. Some results of paper \cite{DH21} indicate that the evolution can be quite
different for the Landau gauge, because of another geometry of the induced electric field. But the general
linear gauge of the time-dependent magnetic field is still an unsolved problem. And what can happen for 
arbitrary superpositions of the Fock states in the case of adiabatic evolution, especially when the magnetic
field changes its sign? We see that the number of unsolved interesting problems is not small.

{
The authors acknowledge the partial support of the Brazilian funding agency 
Conselho Nacional de Desenvolvimento Cient\'{\i}fico e Tecnol\'ogico (CNPq). }

\appendix

\renewcommand{\theequation}{A.\arabic{equation}}
\setcounter{equation}{0}

\section{Integrals with squares of the Laguerre polynomials}
\label{ap-intLag}

Since $\left| \exp\left[ir^2\dot\vep/(2\hbar\vep)  \right] \right|^2 = \exp\left(-Kr^2\right)$, due to identity (\ref{Wr}),
the calculation of the mean values $\langle r^{2j} \rangle$ in the state (\ref{psiFock-t}) can be reduced 
(using the substitution $x= Kr^2$) to the calculation of the integrals
\be
I_j = \int_0^{\infty}  x^{a+j} e^{-x} \left[L_n^{(a)}(x)\right]^2 dx.
\label{I}
\ee
These integrals are absent in the available reference books (such as \cite{Grad}, for example).
However, it is easy to see that 
$I_1 = - \lim_{b\to 1}\partial J(b)/\partial b $, 
where $J(b)$ is the special case of a more general integral 7.414.4 from \cite{Grad}:
\beqn
J(b) &=& \int_0^{\infty}  x^{a} e^{-bx} \left[L_n^{(a)}(x)\right]^2 dx \nonumber
\\ &=& \frac{(2n +a)! (b-1)^{2n}}
{(n!)^2 b^{2n+a+1}} F\left[-n,-n; -2n-a; \frac{b(b-2)}{(b-1)^2}\right].
\eeqn
Here, $F(u,v:w;z)$ is the Gauss hypergeometric function. Using the definition of this function, we can write
\beqn
J(b) &=& b^{-2n-a-1}\left\{ \frac{(n+a)!}{n!}[b(2-b)]^n 
+ (b-1)^2 [b(2-b)]^{n-1} \frac{(n+a +1)!}{(n-1)!} \right. \nonumber \\
&& \left. + (b-1)^4 f(b)\right\},
\label{Jbexp}
\eeqn
where $f(b)$ is some function, which is regular at the point $b=1$. 
Calculating the derivative of function (\ref{Jbexp})
with respect to $b$ and putting $b=1$ in the final expression, we arrive at the formula
\be
I_1 = \frac{(n+a)!}{n!} (2n+a+1).
\label{Ires}
\ee
Its consequence is formula (\ref{meanr2}).
Similarly, 
\be
I_2 =  \lim_{b\to 1}\partial^2 J(b)/\partial b^2  =
\frac{(n+a)!}{n!} \left[6n(n+a+1) +(a+1)(a+2)\right].
\ee

\section{Calculation of coefficients in the expansion (\ref{expan})}
\label{ap-Jac}

The scalar product $\langle \psi_{q_r m}|\Psi_{n_r m}(t) \rangle$ can be calculated with the help of
Equation 7.414.4 from \cite{Grad},
\beqn
I &=& \int_0^{\infty} x^{a} e^{-bx} L_n^{(a)}(\lambda x) L_q^{(a)}(\mu x) dx 
\nonumber \\ &=& 
\frac{(n +q +a)! (b-\lambda)^{n} (b-\mu)^{q}}{n! q! b^{n +q +a+1}} 
F\left[-n,-q; -n -q	-a; \frac{b(b-\lambda - \mu)}{(b-\lambda)(b-\mu)}\right].
\label{defI}
\eeqn
The Gauss hypergeometric function in the right-hand side is a polynomial of the order
$n_{<} = \mbox{min}(n,q)$.  One of possible forms is the expression in terms of the Jacobi
polynomial (see Equation 4.22.1 from \cite{Szego}):
\be
P_n^{(\alpha,\beta)}(x) = \frac{(2n +\alpha +\beta)!}{n! (n +\alpha +\beta)!} \left(\frac{x-1}{2}\right)^n
F\left(-n, -n-\alpha; -2n -\alpha -\beta; \frac{2}{1-x} \right).
\ee
Then, Formula (\ref{defI}) can be written as follows (assuming $q \ge n$),
\be
I = \frac{(q+|m|)! (b - \mu)^{q-n} (\lambda+\mu -b)^n}{q! b^{q+|m|+1}}
P_{n}^{(q-n, |m|)}\left( \frac{b(\lambda+\mu -b) - 2\lambda\mu}{b(b -\lambda-\mu)} \right).
\label{I-Jac}
\ee

In our case, we have $x=r^2$ and $\lambda = |\omega|/\hbar$. Other parameters are as follows,
\[
b= \frac{1}{2\hbar}\left(|\omega| - \frac{i\dot\vep}{\vep}\right) = 
\frac{|\omega| u_{+}e^{i\phi}}{\hbar z}, \quad
\mu = \frac{|\omega|} {\hbar |z|^2},
\qquad z = u_{+}e^{i\phi} + u_{-}e^{-i\phi},
\]
\[
b - \lambda = -\,\frac{|\omega| u_{-}e^{-i\phi}}{\hbar z}, \qquad
b - \mu = \frac{|\omega| u_{-}^* e^{i\phi}}{\hbar z^*}, \qquad
\lambda + \mu - b = \frac{|\omega| u_{+}^* e^{-i\phi}}{\hbar z^*}.
\]
The phase $\phi$ was defined in Equation (\ref{adsol+}).
Consequently, the argument of the Jacobi polynomial in Equation (\ref{I-Jac}) equals
\[
 \frac{b(\lambda+\mu -b) - 2\lambda\mu}{b(b -\lambda-\mu)} = \frac{1- |u_{-}|^2}{1+ |u_{-}|^2}.
\]
Combining all terms in the formulas (\ref{psiFock}), (\ref{psiFock-t}) and (\ref{I-Jac}),
we obtain the following expression for the complex coefficient 
$\langle \psi_{q_r m}|\Psi_{n_r m}(t) \rangle$
with $q_r \ge n_r$:
\beqn
\langle \psi_{q_r m}|\Psi_{n_r m}(t) \rangle &=&
\left[\frac{(q_r +|m|)! n_r!}{(n_r +|m|)! q_r!} \left( \frac{z}{z^*}\right)^{2q_r +|m| +1}
\right]^{1/2} e^{i\chi(t) -i\phi(t)(2n_r +|m| +1)}
\nonumber \\ && \times
\frac{\left(u_{-}^*\right)^{q_r - n_r} \left(u_{+}^*\right)^{n_r}}
{u_{+}^{q_r +|m|+1}}
P_{n_r}^{(q_r-n_r, |m|)}\left( \frac{1- |u_{-}|^2}{1+ |u_{-}|^2} \right).
\eeqn
Similar formulas (in a slightly different context) were found in paper \cite{MMT70}.
All phase factors disappear when one calculates the module squared 
$|\langle \psi_{q_r m}|\Psi_{n_r m}(t) \rangle|^2$. Moreover, this quantity becomes symmetric
with respect to quantum numbers $q_r$ and $n_r$. The result is given by Equation (\ref{probab})
of the main text.

\end{document}